\documentstyle[12pt]{article}
\begin{document}
\title{THE ELUSIVE MONOPOLE}
\author{B.G. Sidharth$^*$\\ Centre for Applicable Mathematics \& Computer Sciences\\
B.M. Birla Science Centre, Hyderabad 500 063 (India)}
\date{}
\maketitle
\footnotetext{E-mail:birlasc@hd1.vsnl.net.in}
\begin{abstract}
In this brief note, we argue that the elusive magnetic monopole arises due to
the strong magnetic effects arising from the non
commutative space time structure at small scales. This would also explain why the monopole
has eluded detection even after seventy years.
\end{abstract}
\section{Introduction}
Ever since Dirac deduced theoretically the existence of the monopole in 1931,
it has eluded physicists\cite{r1}. At the same time the possibility of
realising huge amounts of energy using monopoles has been an exciting prospect.
In 1980 when the fiftieth Anniversary of the monopole was being commemorated, Dirac
himself expressed his belief that the monopole did not exist\cite{r2}. Some
scholars have indeed dismissed the monopole\cite{r3,r4}, while in a model based
on quantized vortices in the hydrodynamical formulation, the monopole field
can be mathematically identified with the momentum vector\cite{r5}. Monopoles
had also been identified with solitons\cite{r6}.\\
In any case, it has been noted that the existence of free monopoles would lead
to an unacceptably high density of the universe\cite{r7}, which in the light of
latest observations of an ever expanding universe\cite{r8,r9} would be difficult
to reconcile.\\
We will now show that monopoles arise due to a non commutative structure of
space time, and this would also provide an explanation for their being undetected.
\section{The Monopole}
The Dirac equation for a spin half particle throws up a complex position coordinate.
Dirac identified the imaginary part with Zitterbewegung effects and argued that
this would be eliminated once it is realized that in Quantum Mechanics, space
time points are not meaningful and that on the contrary averages over intervals
of the order of the Compton scale have to be taken to recover meaningful
physics\cite{r10}. Over the decades the significance of such cut off space time
intervals has been stressed by T.D. Lee and several other
scholars\cite{r11,r12,r13,r14}. Indeed with a minimum cut off length $l$, it was
shown by Snyder\cite{r15} that there would be a non commutative space time
structure, and infact at the Compton scale we would have (Cf. ref.\cite{r14})
\begin{equation}
\left[ x,y\right] = 0 (l^2)\label{e1}
\end{equation}
and similar relations.\\
This feature has been also stressed in recent times by Witten\cite{r16}, who has
pointed out that conventional Quantum Mechanics and Quantum Field Theory are
based on a Minkowskian or Bosonic space time, where as there is a Fermionic,
non commutative space time structure at smaller scales. This is given by equations like (\ref{e1}).\\
We now observe that the non commutative structure given by (\ref{e1}) can be
shown to be associated with a strong magnetic field $B$\cite{r17} given by
$$\frac{\hbar c}{eB} \sim l^2$$
Whence we have
\begin{equation}
\mu = Bl^2 \sim \frac{\hbar c}{e}\label{e2}
\end{equation}
Equation (\ref{e2}) infact defines the magnetic monopole. This means that the
magnetic monopole is thrown up at the Compton scale or, equivalently for masses
near the Planck mass, the Planck scale, due to the non commutative
or Fermionic space structure.\\
Interestingly a similar explanation has been shown to hold for Quarks also, and
it was pointed out that this meant that the Quarks would automatically be confined
and would not be observed as free entities\cite{r18}. For the same reason the
monopole also would not, in this formulation be observed as a free entity.
This could explain why they have eluded detection for some seventy years now.\\
Interestingly an equation like (\ref{e1}) shows that there would be a break
down of parity at the Compton scale. This can be easily verified from the
fact that the bi spinor Dirac wave function has "positive energy" components
$\theta$ and "negative energy" components $\chi$ and that as we approach the
Compton wavelength it is the component $\chi$ which predominates, and also
exhibits the opposite parity\cite{r19}.


\begin{thebibliography}{99}
\bibitem {r1} P.A.M. Dirac, Proc. Roy. Soc., \underline{A 133}, 1931, pp.60 ff.
\bibitem {r2} P.A.M. Dirac, in "Monopoles in Quantum Field Theory", Eds. N.S.
Craigie, P. Goddard and W. Nahm, World Scientific, Singapore, 1982, p.iii.
\bibitem {r3} M. Sachs, Il Nuovo Cimento, Vol.114 B, No.2, February 1999,
p.123-126.
\bibitem {r4} A.J. Staruszkiewicz in "Quantum Coherence and Reality" (Y. Aharanov,
Fetschrift), Eds. J. Anandan and J.L. Sifko, World Scientific, Singapore,
1994, pp.90-94.
\bibitem {r5} B.G. Sidharth, Ind.J.Pure and Appl.Phys., Vol.35, July 1997,
pp.456-471.
\bibitem {r6} D.I. Olive, Nuc.Phys. B (Proc.Suppl.) 46, 1996, pp.1-15.
\bibitem {r7} J.V. Narlikar, "Introduction to Cosmology", Cambridge University
Press, Cambridge, 1993, p.57.
\bibitem {r8} S. Perlmutter, et al, Nature, Vol.391, 1 January 1998, p.51-59.
\bibitem {r9} P. Coles, and  G.F.R. Ellis, "Is the Universe Open
or Closed?", Cambridge University Press, Cambridge, 1997.
\bibitem {r10} P.A.M. Dirac, "The Principles of Quantum Mechanics", Clarendon
Press, Oxford, 1958, pp.4ff, pp.253ff.
\bibitem {r11} L. Bombelli, J. Lee, D. Meyer and R.D. Sorkin, Physical
Review Letters, Vol.59, No.5, August 1987, p.521-524.
\bibitem {r12} T.D. Lee, Physics Letters, Vol.122B, No.3,4, 10March 1983,
p.217-220.
\bibitem {r13} V.G. Kadyshevskii, Translated from Doklady Akademii Nauk
SSSR, Vol.147, No.6, December 1962, p.1336-1339.
\bibitem {r14} B.G. Sidharth, Chaos, Solitons and Fractals, 11(2000), p.1269-1278.
\bibitem {r15} H.S. Snyder, Physical Review, Vol.72, No.1, July 1 1947, p.68-71.
\bibitem {r16} W. Witten, Physics Today, April 1996, pp.24-30.
\bibitem {r17} T. Saito, Gravitation and Cosmology, 6 (2000), No.22, pp.130-136.
\bibitem {r18} B.G. Sidharth, Mod.Phys.Lett.A., Vol.14, No.5, 1999, p.387-389.
\bibitem {r19} J.D. Bjorken and S.D. Drell, "Relativisitic Quantum Mechanics",
Mc-Graw Hill, New York, 1964, p.24.
\end{thebibliography}
\end{document}